# Spin Injection and Inverse Edelstein Effect in the Surface States of Topological Kondo Insulator SmB$_6$

Qi Song[1,2], Jian Mi[1,2], Dan Zhao[3,4], Tang Su[1,2], Wei Yuan[1,2], Wenyu Xing[1,2], Yangyang Chen[1,2], Tianyu Wang[1,2], Tao Wu[3,4,5], Xian Hui Chen[3,4,5,6], X. C. Xie[1,2], Chi Zhang[1,2]*, Jing Shi[7]*, and Wei Han[1,2]*

[1]International Center for Quantum Materials, School of Physics, Peking University, Beijing 100871, China

[2]Collaborative Innovation Center of Quantum Matter, Beijing 100871, P. R. China

[3]Hefei National Laboratory for Physical Science at Microscale and Department of Physics, University of Science and Technology of China, Hefei, Anhui 230026, China

[4]Key Laboratory of Strongly-coupled Quantum Matter Physics, University of Science and Technology of China, Chinese Academy of Sciences, Hefei 230026, China

[5]Collaborative Innovation Center of Advanced Microstructures, Nanjing University, Nanjing 210093, China

[6]High Magnetic Field Laboratory, Chinese Academy of Sciences, Hefei, Anhui 230031, China

[7]Department of Physics and Astronomy, University of California, Riverside, California 92521, USA

*Correspondence to: weihan@pku.edu.cn (W.H.); jing.shi@ucr.edu (J.S.); gwlzhangchi@pku.edu.cn (C.Z.)

There has been considerable interest in exploiting the spin degrees of freedom of electrons for potential information storage and computing technologies. Topological insulators (TI), a class of quantum materials, have special gapless edge/surface states, where the spin polarization of the Dirac fermions is locked to the momentum direction. This spin-momentum locking property gives rise to very interesting spin-dependent physical phenomena such as the Edelstein and inverse Edelstein effects. However, the spin injection in pure surface states of TI is very challenging because of the coexistence of the highly conducting bulk states. Here, we experimentally demonstrate the spin injection and observe the inverse Edelstein effect in the surface states of a topological Kondo insulator, $SmB_6$. At low temperatures when only surface carriers are present, a clear spin signal is observed. Furthermore, the magnetic field angle dependence of the spin signal is consistent with spin-momentum locking property of surface states of $SmB_6$.

**Introduction**

Spintronics aims to use the spin degrees of freedom for information technologies[1-3]. The injection of spin-polarized carriers into two dimensional quantum materials, including graphene and the surface states of topological insulators (TI), is particularly interesting[4,5]. Different from graphene showing weak spin-orbit coupling and long spin lifetimes[6-9], the surface states of TI exhibit very large spin-orbit coupling[10-13]. Even more interestingly, the spin and the momentum directions are strongly coupled to each other in the surface states of TI[4,10-14]. Since the observation of the spin-momentum locking properties with scanning tunneling microscopy and spin- angle-resolved photoemission spectroscopy (spin-APERS)[15,16], a great deal of effort has been made to demonstrate various unique effects associated with this property, such as large spin polarization currents and large spin-orbit torque in the $Bi_2Se_3$-based three dimensional (3D) TI[17-25]. However, a major obstacle to the clean demonstration of the Edelstein/inverse Edelstein effects for the spin-momentum locked surface states is the presence of unavoidable bulk carriers which dominate the conduction in these $Bi_2Se_3$-based 3D TI[19,26]. Recently, $SmB_6$, a Kondo insulator, has been found to be a new type of TI based on transport measurements and angle-resolved photoemission spectroscopy[27-35]. At temperatures below ~ 3 K, the bulk states are insulating, and only surface carriers contribute to the conduction, as demonstrated by the previous surface Hall measurements[30,31].

Here, we report the spin injection into the surface states using the spin pumping and the observation of the inverse Edelstein effect in this topological Kondo insulator (TKI). The temperature and magnetic field angle dependences of the spin voltage are consistent with the spin-momentum locking properties of the surface states, which have been shown to be topological in previous studies[29].

**Results**

**Spin injection into the surface states of $SmB_6$**

The spin injection experiment is performed using $Ni_{80}Fe_{20}$ (Py) as the spin injector, which is deposited onto the (001) surface of the $SmB_6$ single crystals, as shown in Fig. 1a (See methods for details). When the ferromagnetic resonance condition for Py is fulfilled under certain magnetic fields and microwave frequencies, the precessing magnetization launches a spin current which enters the adjacent nonmagnetic $SmB_6$ layer. This technique is called spin pumping, which has been widely used to measure the spin to charge conversion in various materials, including metals, semiconductors, and graphene, etc[36-43]. In our measurements, we use a radio frequency (RF) signal generator to provide the microwave power and standard lock-in technique for better sensitivity and signal to noise ratio (see methods for details). Fig. 1b shows the schematic drawing of energy dispersion relationship of the surface states at the Fermi level for both $\bar{X}$ and $\Gamma$ points. The resistance of the SmB$_6$ device is measured from 300 K to ~ 0.8 K, as shown in Fig. 1c. Clearly, the resistance saturates blow ~ 3 K, which indicates that the surface states are dominant and the bulk states do not contribute to conduction. As the temperature increases, the resistance decreases quite rapidly, due to a large number of the activated bulk carriers as the temperature increases.

Fig. 1d shows the typical magnetic field dependence of the spin voltage measured at 1.7 K with three representative microwave frequencies of 8.3, 9.4, and 10.1 GHz, respectively. We first confirm that the magnetic fields, at which we observe the voltage signals, are the same as the resonance magnetic fields ($H_{res}$) of the Py under the same microwave frequencies (Supplementary Figure 1 and Supplementary Note 1). It is noticed that there are mainly three contributions to the voltages, namely the voltage due to the spin pumping and inverse Edelstein effect ($V_{SP}$), the voltage due to the Seebeck effect from the microwave heating ($V_{SE}$), and the anomalous Hall effect ($V_{AHE}$) of the Py. Due to their different symmetries as a function of the magnetic field, we can obtain the voltage amplitudes of all these three contributions by fitting the magnetic field dependence of the voltage with the following equation (Supplementary Figure 2 and Supplementary Note 2).

$$V(H) = V_S \frac{(\Delta H)^2}{(H - H_{res})^2 + (\Delta H)^2} + V_{AS} \frac{-2\Delta H (H - H_{res})}{(H - H_{res})^2 + (\Delta H)^2} \qquad (1)$$

where $V_S$ and $V_{AS}$ are the voltage amplitudes for the symmetric and antisymmetric Lorentzian shapes, respectively, and $\Delta H$ is the half-line width. The $V_{SP}$ exhibits a positive sign for positive magnetic fields and the sign of the spin-to-charge conversion in the surface states of the SmB$_6$ is theoretically expected from the counter-clockwise spin textures for the electron band of the topological surface states[18,42]. The counter-clockwise spin textures have been shown by both spin-APERS measurements and DFT calculations[34,44]. After the determination of $H_{res}$ and $\Delta H$ for all applied microwave frequencies, we obtain the effective magnetization ($M_{eff}$) and the Gilbert damping constant for the Py layer. Our results show that $M_{eff}$ is $781 \pm 16 \, emu \cdot cm^{-3}$, which is obtained using the Kittel formula shown below[45]:

$$f_{res} = (\frac{\gamma}{2\pi})[H_{res}(H_{res} + 4\pi M_{eff})]^{1/2} \quad (2)$$

where $\gamma$ is the geomagnetic ratio. From the slope of the linearly fitted curve of the half-line width vs microwave frequency at 1.7 K, we calculate the Gilbert damping constant of the Py on SmB$_6$ to be $0.0166 \pm 0.0006$ (Supplementary Figure 3).

The microwave power dependence of the spin voltage is shown in Fig. 2a measured at 1.7 K and with the microwave frequency of 10.1 GHz. The measured resonance peak increases as the microwave power increases. Following the same fitting procedure (Supplementary Note 2), we obtain the power dependence of $V_{SP}$ and $V_{SE}$. Both $V_{SP}$ and $V_{SE}$ show a linear relationship with the microwave power, as shown in Fig. 2b and 2c.

**Temperature dependence of the spin voltage**

As mentioned earlier, the surface states of SmB$_6$ dominate the transport as the bulk carriers freeze out below ~ 3 K; above ~ 3 K, the contribution from the bulk states is thermally activated.

When a spin current enters the spin-momentum locked surface states, an electric field is resulted due to the inverse Edelstein effect which is measured as a spin voltage. To investigate how the spin voltage evolves as the surface states emerge and become dominant, we perform the measurements from ∼ 0.8 to 10 K. Below ~ 0.8 K, it is difficult to stabilize the temperature due to the microwave heating. Fig. 3a shows the typical measurements of the voltage as a function of the magnetic field with the microwave power of 100 mW and frequency of 10.1 GHz at 0.84, 1.66, 2.1, 2.3 and 10 K, respectively. At 0.8 K, when only spin-momentum locked surface states exist, the spin signal is ~ 42 nV. This value is relatively small compared to previous studies on $Bi_{1.5}Sb_{0.5}Te_{1.7}Se_{1.3}$ and α-Sn [18,42], which could be related to the spin pumping efficiency and/or the spin-to-charge conversion efficiency and needs further studies (Supplementary Note 3). The spin voltage steadily decreases as the temperature increases, and when the temperature reaches 10 K, no voltage can be detected. The resistance of the $SmB_6$ as a function of the temperature is shown in Fig. 1b, indicating that the bulk states start to contribute to the total conductance between 2 and 3 K. From 3 to 5 K, the conduction due to the bulk carriers quickly increases, resulting in a 100-fold decrease in the total resistance. This feature is consistent with the previous surface conductance and Hall measurements, indicating the nearly pure surface states contributing to the conduction[30,31]. The temperature dependence of the $V_{SP}$ is summarized in Fig. 3c. $V_{SP}$ shows little temperature dependence below ~ 2.2 K. At temperatures above ~ 2.2 K, $V_{SP}$ steadily decreases as the temperature increases. The temperature dependences of both $V_{SP}$ and the resistance strongly support that the spin signal originates from the spin-momentum locked surface states. When the spin polarization is generated in the surface states, an in-plane electrical voltage is produced in the direction perpendicular to the spin directions, due to the inverse Edelstein effect. As the temperature further increases, more bulk carriers are activated, and the spin voltage is greatly suppressed. This is very interesting, for the bulk states should have strong-spin orbit coupling as well and therefore ordinary inverse spin Hall effect from bulk states could give rise to a finite voltage. However, we do not observe any voltage signal at high temperatures.

**Magnetic field angle dependence of the spin voltage**

To further confirm the spin injection and detection in the surface states of the TKI, $SmB_6$, we study the in-plane and out-of-plane spin polarization injection by changing the magnetic field direction. Fig. 4a shows the typical results of the magnetic field dependent voltages at 1.7 K with a microwave power of 200 mW and frequency of 10.1 GHz for the angles between the magnetic field and the Py electrode (shown in the inset figure), $\theta_H$, equal to 0°, 63°, 76°, 83°, and 86.5°. As $\theta_H$ increases, the resonance magnetic field increases accordingly, and in the meantime, the spin signal shows a decrease. At 86.5°, the spin-dependent voltage becomes vanishingly small. The $H_{res}$ and $\Delta H$ as a function of $\theta_H$ are shown in Fig. 4b and 4c, which are consistent with the previous measurement of the ferromagnetic resonance of Py under different magnetic field directions[37,46]. This further confirms that the measured spin voltage indeed arises from the precession of the Py magnetization.

**Discussion**

It is particularly interesting that only in-plane spin polarization injection generates an electric field, whereas the out-of-plane spin polarization injection does not show this effect. This observation could be attributed to the spin-momentum locking properties of the surface states of the TKI, as illustrated in Fig. 5a and 5b. For the in-plane spin polarization injection along the $x$ direction, the Fermi surface shifts along the y direction, and $\Delta k_y$ indicates the total shift due to the spin injection and the inverse Edelstein effect, as shown in Fig. 5a. On the other hand, for the out-of-plane spin polarization injection, there is no net effect of spin injection since the spins of the surface states lie in-plane and are locked perpendicular to the momentum directions, as shown in Fig. 5b. Finally, we calculate the Py magnetization angle, $\theta_M$, from the $\theta_H$ dependence of the resonance magnetic field (Supplementary Figure 4) based on the 0 and 90 degrees data and the following equation[37].

$$2H_{res}\sin(\theta_H - \theta_M) - 4\pi M_S \sin(2\theta_M) = 0 \qquad (3)$$

where $M_S$ is the saturated magnetization. It is clearly seen that $V_{SP}$ almost vanishes as $\theta_M$ approaches 90 degrees (Fig. 5c), which is also consistent with the spin-momentum locking properties of the surface states of the TKI, as discussed above and illustrated in Fig. 5a and 5b. The complete understanding of the $V_{SP}$ as a function of the $\theta_M$ needs future theoretical studies to quantitatively calculate how much the Fermi surface shift as a result of the inverse Edelstein effect of the spin polarization injection (Supplementary Figure 5 and Note 4).

Our experimental results strongly support the demonstration of spin injection and the observation of the inverse Edelstein effect in the surface states of $SmB_6$. The temperature and magnetization angle dependence, as well as the sign of the spin-to-charge conversion are well consistent with spin-momentum locking properties of the surface states, which have been shown to be topological with the counter-clockwise spin textures for the electron bands in previous studies[34,44]. Since the detailed spin textures of the Rashba surface states have not been reported yet, it is premature to exclude any contribution from the Rashba-split surface states at the current stage. To fully understand this, further studies, including the detailed spin textures from spin-APERS measurements of the Rashba surface states and the quantitative theoretical calculations of the contributions from topological and Rashba surface states, are needed. Our observation could lead to future studies of the role of strong correlation in TKIs for spintronics and highly efficient spin current generation in the surface states of TIs via the materials design and engineering.

**Methods**

**Materials growth**

High-quality single crystalline $SmB_6$ samples are grown using the conventional Al-flux method. A mixture consisting of a Sm chunk (purity: 99.9%), Boron (purity: 99.99%) and Al powders (purity: 99.99%) with a ratio of 1:6:400 is heated at high temperatures in the circumstance with continuously flowing Ar gas to form $SmB_6$ single crystals. Then the $SmB_6$ samples are put into diluted $HNO_3$ acid to remove the residual aluminum flux.

We choose the samples with large rectangular crystals of millimeters size and large (001) facet for spin injection experiment. A 20 nm thick Py electrode is deposited on the (001) surface of the $SmB_6$ single crystal by RF magnetron sputtering with a growth rate of 0.02 Å/s. To prevent the oxidation of Py, a capping layer of 3 nm Al is deposited *in-situ* prior to taking the samples out.

**Device fabrication**

A shadow mask technique (size: ~ 0.9 × 3 $mm^2$) is used to define the shape and position of the ferromagnetic electrode (Py/Al) on the (001) surface of the $SmB_6$ crystal (size: ~ 1 × 5 $mm^2$, thickness: ~ 0.5 mm). Al wires are used to contact the two ends of $SmB_6$ sample for the electrical voltage measurement.

**Device measurement**

The spin injection is performed using the spin pumping method and the spins are detected via the inverse Edelstein effect of the surface states of $SmB_6$. The microwave power is supplied by a signal generator (Anritsu LTD. MG3690C) modulated with a digital lock-in amplifier (NF Co. LI5640) with the frequency of 373 Hz to enhance the sensitivity and signal-to-noise ratio. The spin pumping measurement is performed by precessing the Py magnetization around its resonance conditions with a coplanar waveguide from 10 to ~ 0.8 K in a Janis He-3 system. The resistance of the $SmB_6$ single crystal is measured using Keitheley K2400 and K2002 in Quantum Design Physical Properties Measurement System (PPMS) from 300 to 10 K and in a Janis He-3 system from 10 to ~ 0.8 K.

**Data Availability** The authors declare that the data supporting the findings of this study are available within the paper and its supplementary information files.

**References:**

1       Wolf, S. A. *et al.* Spintronics: A Spin-Based Electronics Vision for the Future. *Science* **294**, 1488-1495, (2001).
2       Zutic, I., Fabian, J. & Das Sarma, S. Spintronics: Fundamentals and applications. *Rev. Mod. Phys* **76**, 323-410, (2004).
3       Bader, S. D. & Parkin, S. S. P. Spintronics. *Annu. Rev. Cond. Mat.* **1**, 71-88, (2010).
4       Pesin, D. & MacDonald, A. H. Spintronics and pseudospintronics in graphene and topological insulators. *Nat. Mater.* **11**, 409-416, (2012).
5       Han, W. Perspectives for spintronics in 2D materials. *APL Mater.* **4**, 032401, (2016).
6       Tombros, N., Jozsa, C., Popinciuc, M., Jonkman, H. T. & van Wees, B. J. Electronic spin transport and spin precession in single graphene layers at room temperature. *Nature* **448**, 571-574, (2007).
7       Han, W., Kawakami, R. K., Gmitra, M. & Fabian, J. Graphene spintronics. *Nat. Nanotech.* **9**, 794-807, (2014).
8       Stephan, R. *et al.* Graphene spintronics: the European Flagship perspective. *2D Materials* **2**, 030202, (2015).
9       Dlubak, B. *et al.* Highly efficient spin transport in epitaxial graphene on SiC. *Nat. Phys.* **8**, 557-561, (2012).
10      Qi, X.-L. & Zhang, S.-C. Topological insulators and superconductors. *Rev. Mod. Phys.* **83**, 1057-1110, (2011).

11	Hasan, M. Z. & Kane, C. L. Colloquium: Topological insulators. *Rev. Mod. Phys.* **82**, 3045-3067, (2010).
12	Moore, J. E. The birth of topological insulators. *Nature* **464**, 194-198, (2010).
13	Ando, Y. Topological Insulator Materials. *J. Phys. Soc. Jpn.* **82**, 102001, (2013).
14	Yazyev, O. V., Moore, J. E. & Louie, S. G. Spin Polarization and Transport of Surface States in the Topological Insulators Bi2Se3 and Bi2Te3 from First Principles. *Phys. Rev. Lett.* **105**, 266806, (2010).
15	Hsieh, D. *et al.* A tunable topological insulator in the spin helical Dirac transport regime. *Nature* **460**, 1101-1105, (2009).
16	Roushan, P. *et al.* Topological surface states protected from backscattering by chiral spin texture. *Nature* **460**, 1106-1109, (2009).
17	Li, C. H. *et al.* Electrical detection of charge-current-induced spin polarization due to spin-momentum locking in Bi2Se3. *Nat. Nanotechnol.* **9**, 218-224, (2014).
18	Shiomi, Y. *et al.* Spin-Electricity Conversion Induced by Spin Injection into Topological Insulators. *Phys. Rev. Lett.* **113**, 196601, (2014).
19	Mellnik, A. R. *et al.* Spin-transfer torque generated by a topological insulator. *Nature* **511**, 449-451, (2014).
20	Fan, Y. *et al.* Magnetization switching through giant spin–orbit torque in a magnetically doped topological insulator heterostructure. *Nat. Mater.* **13**, 699-704, (2014).
21	Tang, J. *et al.* Electrical Detection of Spin-Polarized Surface States Conduction in (Bi0.53Sb0.47)2Te3 Topological Insulator. *Nano Letters* **14**, 5423-5429, (2014).
22	Ando, Y. *et al.* Electrical Detection of the Spin Polarization Due to Charge Flow in the Surface State of the Topological Insulator Bi1.5Sb0.5Te1.7Se1.3. *Nano Letters* **14**, 6226-6230, (2014).
23	Liu, L. *et al.* Spin-polarized tunneling study of spin-momentum locking in topological insulators. *Phys. Rev. B* **91**, 235437, (2015).
24	Wang, Y. *et al.* Topological Surface States Originated Spin-Orbit Torques in Bi2Se3. *Phys. Rev. Lett.* **114**, 257202, (2015).
25	Lee, J. S., Richardella, A., Hickey, D. R., Mkhoyan, K. A. & Samarth, N. Mapping the chemical potential dependence of current-induced spin polarization in a topological insulator. *Phys. Rev. B* **92**, 155312, (2015).
26	Jiang, Z. *et al.* Enhanced spin Seebeck effect signal due to spin-momentum locked topological surface states. *Nat. Commun.* **7**, 11458 (2016).
27	Dzero, M., Sun, K., Galitski, V. & Coleman, P. Topological Kondo Insulators. *Phys. Rev. Lett.* **104**, 106408, (2010).
28	Lu, F., Zhao, J., Weng, H., Fang, Z. & Dai, X. Correlated Topological Insulators with Mixed Valence. *Phys. Rev. Lett.* **110**, 096401, (2013).
29	Dzero, M., Xia, J., Galitski, V. & Coleman, P. Topological Kondo Insulators. *Annu. Rev. Condens. Matter Phys.* **7**, 249-280, (2016).
30	Kim, D. J., Xia, J. & Fisk, Z. Topological surface state in the Kondo insulator samarium hexaboride. *Nat. Mater.* **13**, 466-470, (2014).
31	Kim, D. J. *et al.* Surface Hall Effect and Nonlocal Transport in SmB6: Evidence for Surface Conduction. *Scientific Reports* **3**, 3150, (2013).
32	Li, G. *et al.* Two-dimensional Fermi surfaces in Kondo insulator SmB6. *Science* **346**, 1208-1212, (2014).

33    Jiang, J. *et al.* Observation of possible topological in-gap surface states in the Kondo insulator SmB6 by photoemission. *Nat. Commun.* **4**, 3010 (2013).
34    Xu, N. *et al.* Direct observation of the spin texture in SmB6 as evidence of the topological Kondo insulator. *Nat. Commun.* **5**, 4566, (2014).
35    Syers, P., Kim, D., Fuhrer, M. S. & Paglione, J. Tuning Bulk and Surface Conduction in the Proposed Topological Kondo Insulator SmB6. *Phys. Rev. Lett.* **114**, 096601, (2015).
36    Mosendz, O. *et al.* Quantifying Spin Hall Angles from Spin Pumping: Experiments and Theory. *Phys. Rev. Lett.* **104**, 046601, (2010).
37    Ando, K. *et al.* Inverse spin-Hall effect induced by spin pumping in metallic system. *J. Appl. Phys.* **109**, 103913, (2011).
38    Ando, K. *et al.* Electrically tunable spin injector free from the impedance mismatch problem. *Nat Mater* **10**, 655-659, (2011).
39    Czeschka, F. D. *et al.* Scaling Behavior of the Spin Pumping Effect in Ferromagnet-Platinum Bilayers. *Phys. Rev. Lett.* **107**, 046601, (2011).
40    Ando, K. & Saitoh, E. Observation of the inverse spin Hall effect in silicon. *Nat Commun* **3**, 629, (2012).
41    Zhang, W. *et al.* Spin Hall Effects in Metallic Antiferromagnets. *Phys. Rev. Lett.* **113**, 196602, (2014).
42    Rojas-Sánchez, J. C. *et al.* Spin to Charge Conversion at Room Temperature by Spin Pumping into a New Type of Topological Insulator: α-Sn Films. *Phys. Rev. Lett.* **116**, 096602, (2016).
43    Dushenko, S. *et al.* Gate-Tunable Spin-Charge Conversion and the Role of Spin-Orbit Interaction in Graphene. *Phys. Rev. Lett.* **116**, 166102, (2016).
44    Yu, R., Weng, H., Fang, Z. & Dai, X. Pseudo-Spin, Real-Spin and Spin Polarization of Photo-emitted Electrons. *arXiv:1603.09677* (2016).
45    Kittel, C. On the Theory of Ferromagnetic Resonance Absorption. *Phys. Rev.* **73**, 155-161, (1948).
46    Luo, C. *et al.* Enhancement of magnetization damping coefficient of permalloy thin films with dilute Nd dopants. *Phys. Rev. B* **89**, 184412, (2014).

**End Notes**

**Acknowledgements**

We acknowledge the fruitful discussions with Professor Fa Wang and the financial support from National Basic Research Programs of China (973 program Grant Nos. 2014CB920902, 2015CB921104, and 2013CB921903) and National Natural Science Foundation of China (NSFC Grant Nos. 11574006 and 11374020). D.Z., T.W. and X.H.C. acknowledge the financial support from the Strategic Priority Research Program of the Chinese Academy of Sciences (Grant No.

XDB04040100). T.W. acknowledges Recruitment Program of Global Experts and CAS Hundred Talent Program. J.S. acknowledges the support by the DOE BES Award No. DEFG02-07ER46351. W.H. acknowledges the support by the 1000 Talents Program for Young Scientists of China.

**Author contributions**

W.H. proposed and designed the experiment. Q.S. did the Py growth and fabricated the devices. Q.S. and J.M. performed the electrical measurements. J.M. and C.Z. developed the microwave techniques in the He3 refrigerator used for the measurements below 10 K in Prof. Zhang's group. Q.S. and W.H. analyzed the data. D.Z., T.W., and X.H.C. provided the single crystalline $SmB_6$ sample. W.H. wrote the manuscript. All authors commented on the manuscript and contributed to its final version.

**Competing financial interests**

The authors declare no competing financial interests.

**Figure Legends**

**Figure 1 | Spin injection into the surface states of $SmB_6$. a,** Schematic drawing of device structure and the spin injection and inverse Edelstein effect measurements. **b**, Schematic drawing of the spin-momentum locking properties of the topological surface states at the $\bar{X}$ and $\Gamma$ points based on previous photoemission spectroscopy measurements and DFT calculations[34,44]. **c**, the resistance of the $SmB_6$ as a function of the temperature. **d**, Typical magnetic field dependence of

the voltage with various GHz microwave frequencies. The power of the microwave is 100 mW and the temperature is 1.7 K. Inset, the resonance frequency ($f$) as a function of the resonance magnetic field ($H_{res}$). The solid line is a fitted curve based on the Kittel formula, equation (2) in the main text.

**Figure 2 | Microwave power dependence of the spin injection into the surface states of SmB$_6$. a,** Magnetic field dependence of the voltage measured at the temperature of 1.7 K and with the microwave frequency of 10.1 GHz and power of 15.8, 50.1, 89.1, 141, and 224 mW, respectively. **b-c,** Microwave power dependence of the measured voltage due to spin pumping and inverse Edelstein effect ($V_{SP}$ in **b**) and the voltage that is related to the Seebeck effect ($V_{SE}$ in **c**).

**Figure 3 | Temperature dependence of the spin injection into the surface states of SmB$_6$. a,** Magnetic field dependence of the voltage measured for the temperatures of 0.84, 1.66, 2.1, 2.3, and 10 K, respectively. The measurement is performed with a microwave power 100 mW and frequency of 10.1 GHz. **b**, The resistance of the SmB$_6$ as a function of the temperature from 10 to ~ 0.8 K. **c**, Temperature dependence of $V_{SP}$.

**Figure 4 | The measured voltage as a function of the magnetic field angle. a**, Magnetic field dependence of the voltage measured at 1.7 K for $\theta_H$ = 0°, 63°, 76°, 83°, and 86.5°, respectively. Inset: the schematic illustration of the coordinate system for magnetic field angle. **b-c**, the resonance magnetic field and half-line width as a function of $\theta_H$.

**Figure 5 | Magnetization angle dependence of the voltage due to spin pumping and inverse Edelstein effect. a-b,** Schematic drawings for the in-plane (a) and out-of-plane (**b**) spin polarization injection into the surface states of $SmB_6$. The in-plane spin polarization injection leads to the generation of the in-plane electric field due to inverse Edelstein effect, while the out-of-plane spin injection is forbidden due to the spin-momentum locking properties of the surface states of $SmB_6$, which have been shown to be topological in previous studies. **c,** $V_{SP}$ as a function of $\theta_M$. Inset: the schematic illustration of the coordinate system for $\theta_M$.

Figure 1

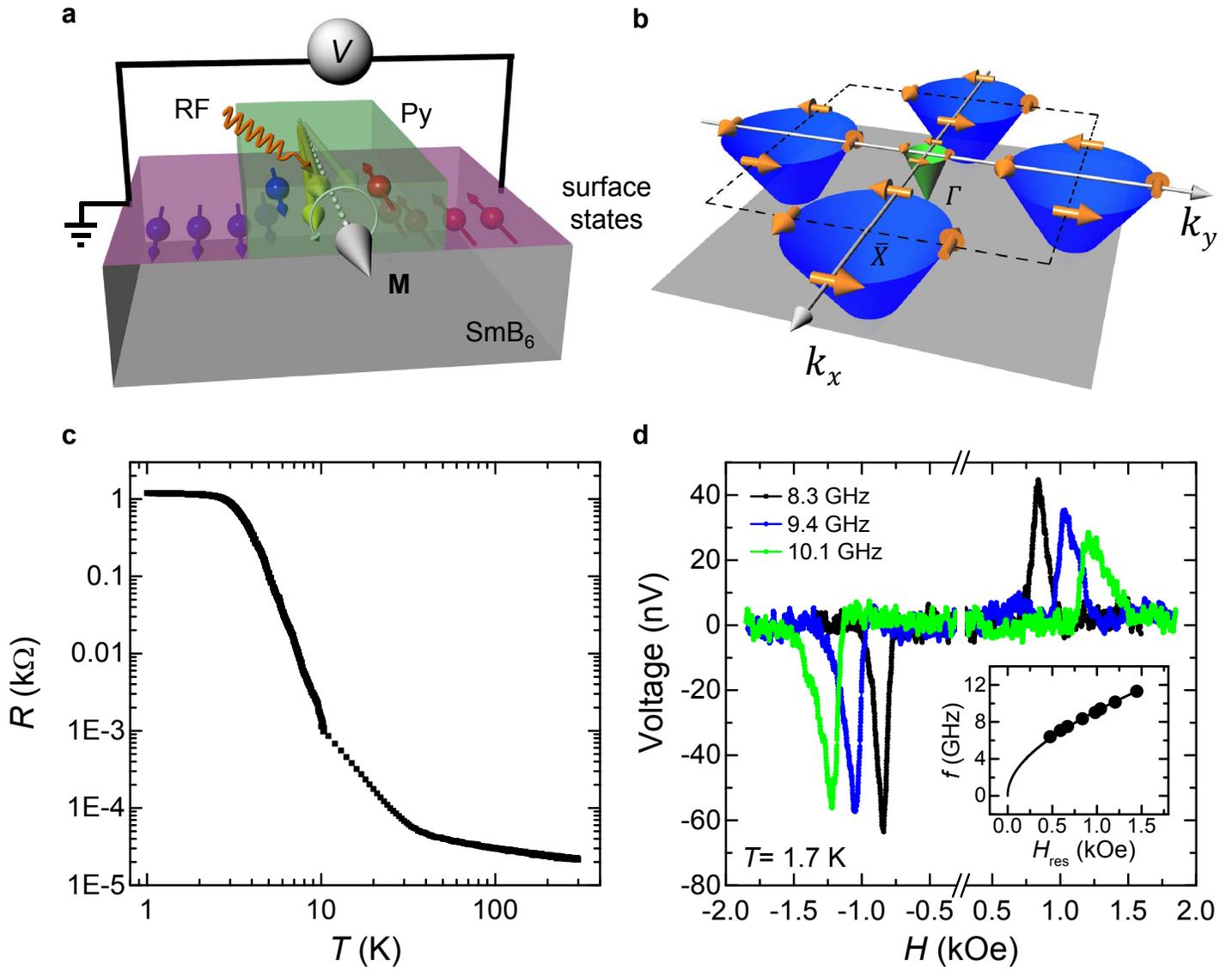

Figure 2

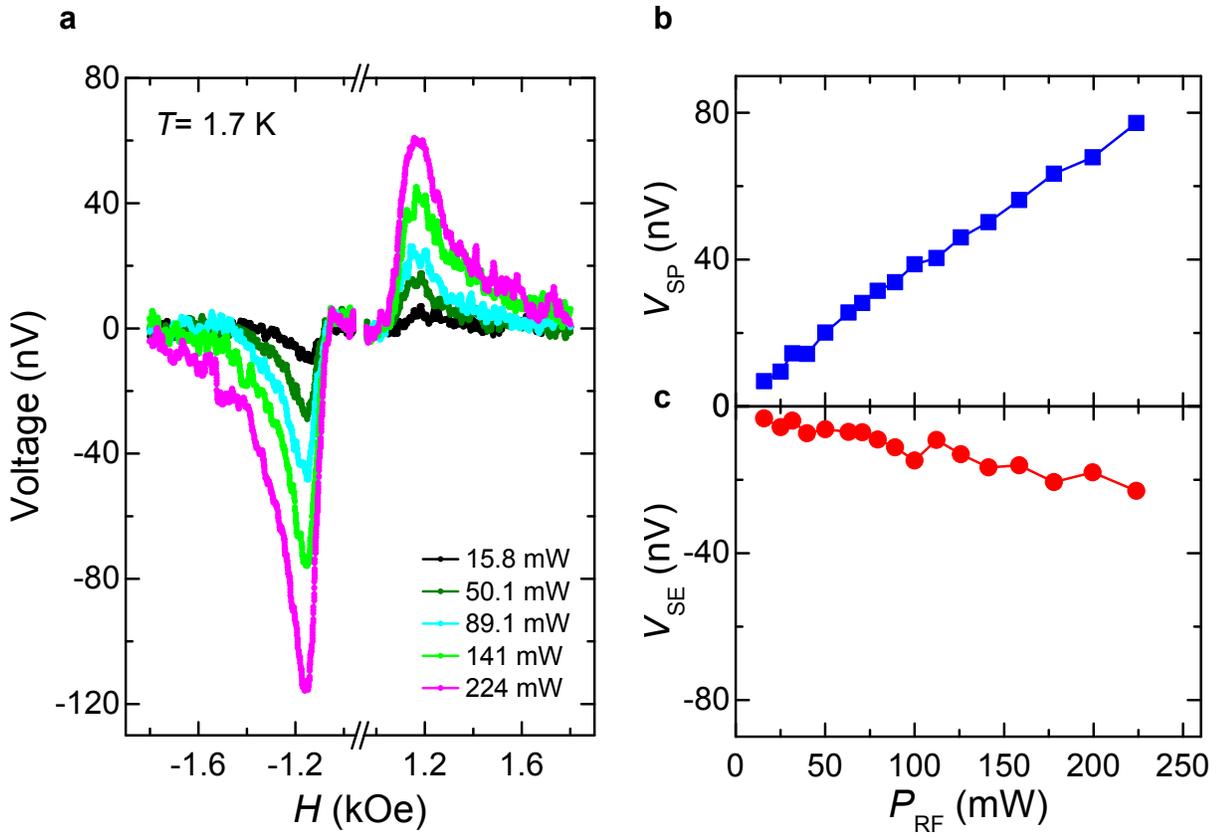

Figure 3

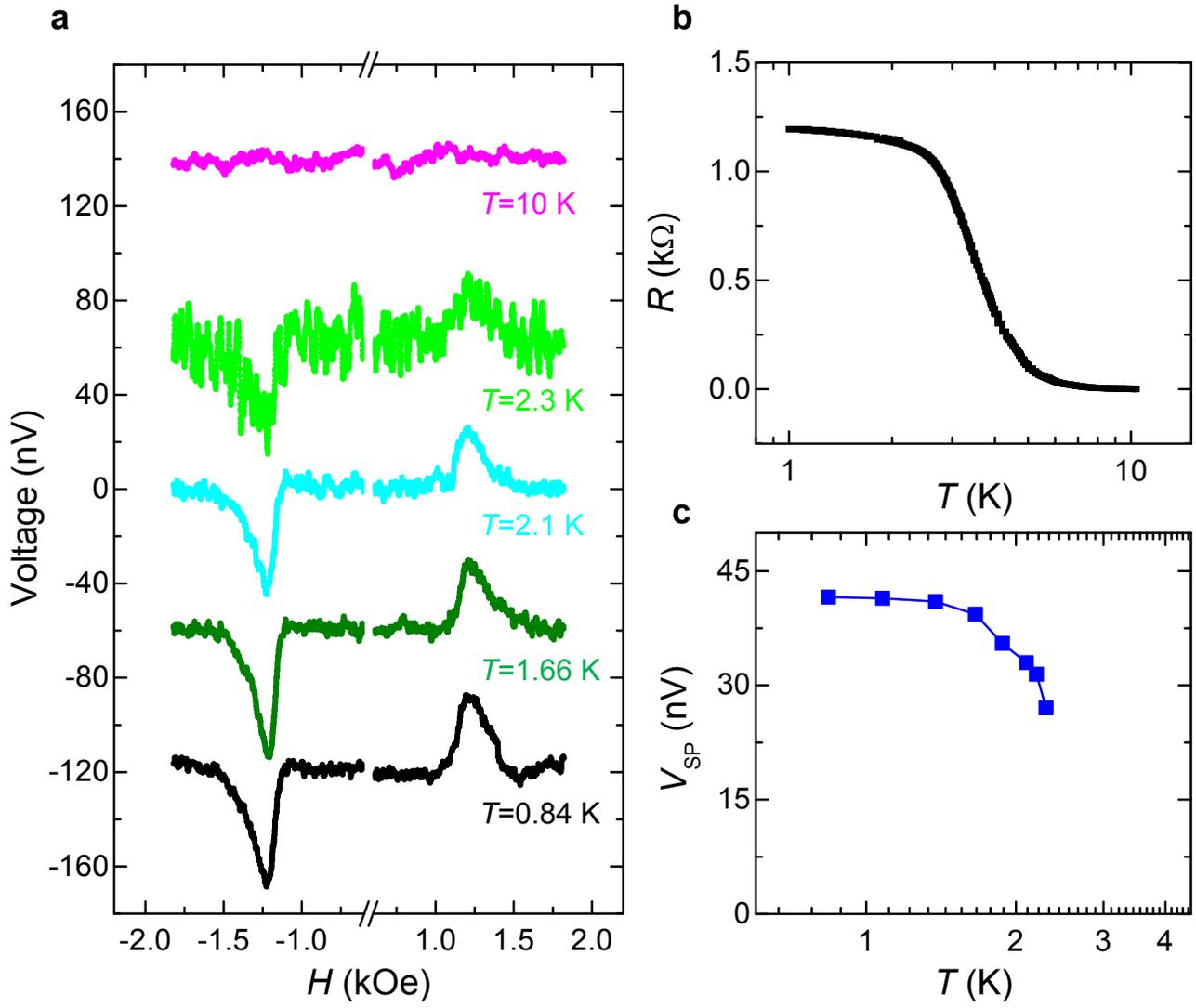

Figure 4

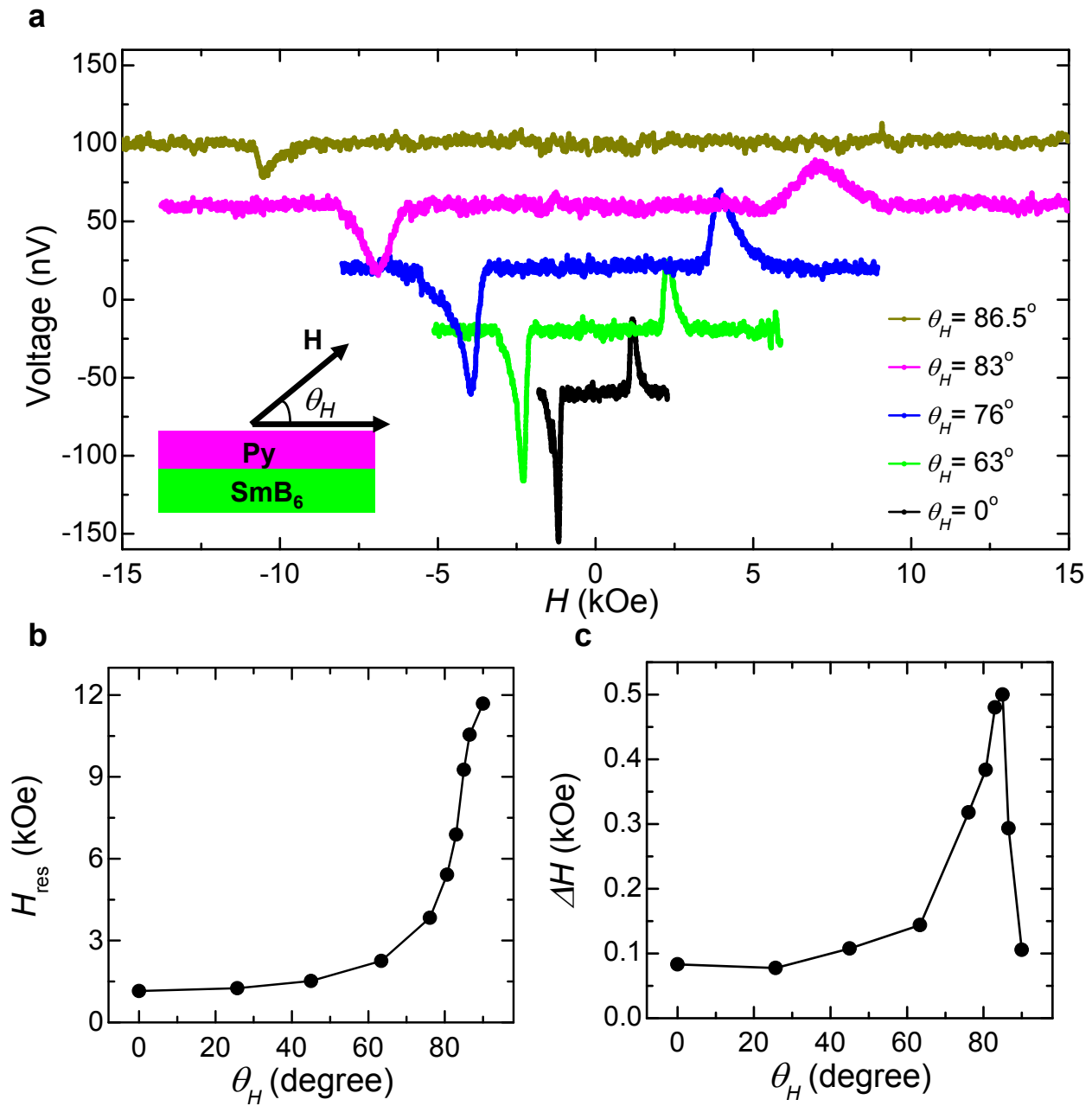

Figure 5

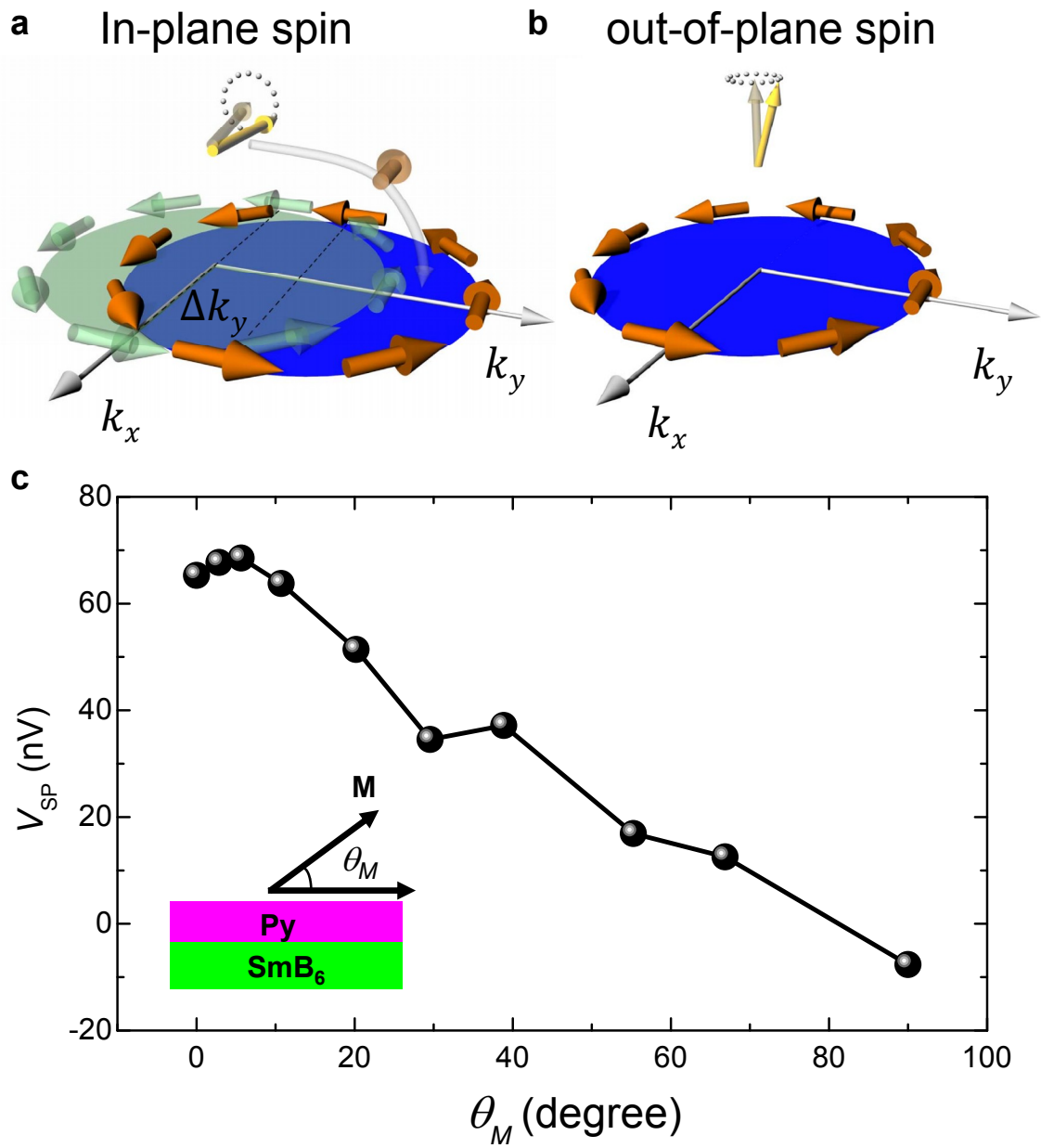